\documentclass[reprint,amsmath,amssymb,aps,prb]{revtex4-1}
\usepackage{graphicx,color,epsfig,rotate}
\usepackage{times,xspace}
\usepackage{amsbsy,amssymb,amsmath,bm}
\usepackage{dcolumn}
\usepackage{bm}

\begin{document}

\title{Multiferroicity with coexisting isotropic and anisotropic spins in Ca$_{3}$Co$_{2-x}$Mn$_{x}$O$_{6}$}

\author{Jae~Wook~Kim$^{1,2}$, Y.~Kamiya$^{3}$, E.~D.~Mun$^{1}$, M.~Jaime$^{1}$, N.~Harrison$^{1}$, J.~D.~Thompson$^{4}$, V.~Kiryukhin$^{5}$, H.~T.~Yi$^{5}$, Y.~S.~Oh$^{5}$, S.-W.~Cheong$^{5}$, C.~D.~Batista$^{3}$, and V.~S.~Zapf$^{1}$}

\affiliation{$^1$National High Magnetic Field Laboratory (NHMFL), Materials Physics and Applications (MPA) - Condensed Matter and Magnet Science (CMMS) group, Los Alamos National Laboratory (LANL), Los Alamos NM 87545, USA}
\affiliation{$^2$Lujan Center for Neutron Scattering, LANL, Los Alamos, NM 87545, USA}
\affiliation{$^3$Theoretical Division, T-4 and Center for Nonlinear Studies, LANL, Los Alamos, NM 87545, USA}
\affiliation{$^4$MPA-CMMS, LANL, Los Alamos NM 87545, USA}
\affiliation{$^5$Rutgers Center for Emergent Materials and Department of Physics and Astronomy, Piscataway, NJ 08854, USA}

\date{\today} 

\pacs{75.85.+t, 75.30.Cr, 75.30.Kz}

\begin{abstract}
We study magnetic and multiferroic behavior in Ca$_3$Co$_{2-x}$Mn$_{x}$O$_6$ ($x \sim$~0.97) by high-field measurements of magnetization ($M$), magnetostriction ($L$($H$)/$L$), electric polarization ($P$), and magnetocaloric effect.
This study also gives insight into the zero and low magnetic field magnetic structure and magnetoelectric coupling mechanisms.
We measured $M$ and $\Delta$$L$/$L$ up to pulsed magnetic fields of 92~T, and determined the saturation moment and field.
On the controversial topic of the spin states of Co$^{2+}$ and Mn$^{4+}$ ions, we find evidence for $S$~=~3/2 spins for both ions with no magnetic field-induced spin-state crossovers.
Our data also indicate that Mn$^{4+}$ spins are quasi-isotropic and develop components in the $ab$-plane in applied magnetic fields of 10~T.
These spins cant until saturation at 85~T whereas the Ising Co$^{2+}$ spins saturate by 25~T.
Furthermore, our results imply that mechanism for suppression of electric polarization with magnetic fields near 10~T is flopping of the Mn$^{4+}$ spins into the $ab$-plane, indicating that appropriate models must include the coexistence of Ising and quasi-isotropic spins.
\end{abstract}
\maketitle

Multiferroic materials exhibit at least two simultaneous long-range orders such as (anti-)ferromagnetism, ferroelectricity, and ferroelasticity \cite{Spaldin05}.
Coupling between magnetic and ferroelectric order parameters leads to magnetoelectric (ME) effects that can be exploited for developing novel functional materials \cite{Scott07,Chu08}.
The microscopic origin of ME coupling in most multiferroics \cite{Cheong07,Kimura07,Khomskii09,Kamiya12a} is thought to be ionic displacements that are sensitive to magnetic order \cite{Hill00}, and/or electronic charge redistribution \cite{Kamiya12a,Jia07,Arima07,Bulaevski08}.
In magnetically-induced multiferroics, ME coupling hinges on magnetic orderings that spontaneously break the spatial-inversion symmetry (SIS), thereby allowing a net electric polarization.
Unfortunately, most SIS-breaking spin structures have little or no net magnetization \cite{Kimura03,Goto04,Hur04,Lawes05,Katsura05,Cheong07,Khomskii09,Kenzelmann07,Arima07,Kimura07,Zapf10} that is coupled to $P$, which limits their eventual usefulness.
In spiral magnets, for example, transverse components of the spins couple to $P$ and the longitudinal $M$ is insensitive to $P$.
Thus, there is an effort in the multiferroics community to find new bulk compounds in which a net $M$ and net $P$ are coupled \cite{Coupling} ideally with hysteresis, and to understand the coupling mechanism. 
Here we study Ca$_3$Co$_{2-x}$Mn$_x$O$_6$ (CCMO) with $x \sim$~0.96~$-$~0.97 \cite{Choi08,Jo09}, which shows  net hysteretic $M$ and $P$ along the same axis. Our goal is to understand the origin of the magnetic
order and the ME coupling in this compound."

The magnetic ordering of CCMO, found from neutron powder diffraction (NPD) measurements \cite{Choi08,Jo09,Kiryukhin09}, is an $\uparrow \uparrow \downarrow \downarrow$ collinear structure of the alternating Co$^{2+}$ and Mn$^{4+}$ spins along chains in the crystallographic $c$-axis at zero magnetic field.
This spin ordering combined with the alternating ionic ordering breaks SIS and thus allows a net $P$, which is observed below magnetic ordering temperature $T_{N}$~=~15~K and $H$~$<$~10~T.
These $c$-axis chains in turn form a hexagonal lattice in the $ab$-plane (see Supplementary Information (SI), Fig.~S1(a)) that likely creates significant frustration.
Similar $\uparrow \uparrow \downarrow \downarrow$ ordering with net hysteretic $M$ coupled to $P$ has also been observed in Lu$_2$MnCoO$_6$ with magnetic ordering temperature $T_c$~=~43~K and $H \leq$~15~T, although in that compound the Co$^{2+}$-Mn$^{4+}$ chains are arranged in a rectangular, rather than hexagonal, configuration in the $ab$-plane \cite{Yanes-Vilar11}.

It has been proposed by several groups \cite{Choi08,Jo09,Lancaster09,Yao09,Flint10} that the magnetic behavior of CCMO results from frustration between nearest and next-nearest-neighbor exchange interactions along the $c$-axis chains.
Several groups mention the axial next-nearest neighbor Ising (ANNNI) model \cite{Bak82,Selke88}, in which frustration on chains of Ising spins creates cascades of different magnetic phases in response to small changes in external parameters.
A hallmark of ANNNI physics is long-wavelength incommensurate modulations of the Ising spins along the chains with temperature ($T$)-dependent wavelengths.
This behavior was observed in the isostructrual compound Ca$_3$Co$_{2}$O$_6$ \cite{Agrestini08,Moyoshi11} and a variant of the ANNNI model has been proposed for Ca$_3$Co$_{2}$O$_6$ \cite{Kamiya12b}.
In this model, frustration between spins on different chains in the $ab$-plane can be mapped onto an effective single chain model with up to third-nearest-neighbor interactions.
Both the ANNNI model and related model for Ca$_3$Co$_{2}$O$_6$ exhibit a transition to commensurate order at a lower temperature with the $\uparrow \uparrow \downarrow \downarrow$ ground state ordering for a certain range of exchange parameters \cite{Fisher80,Kamiya12b}.

\begin{figure}
\hspace*{-0.5cm}
\includegraphics[angle=0,width=0.85\hsize]{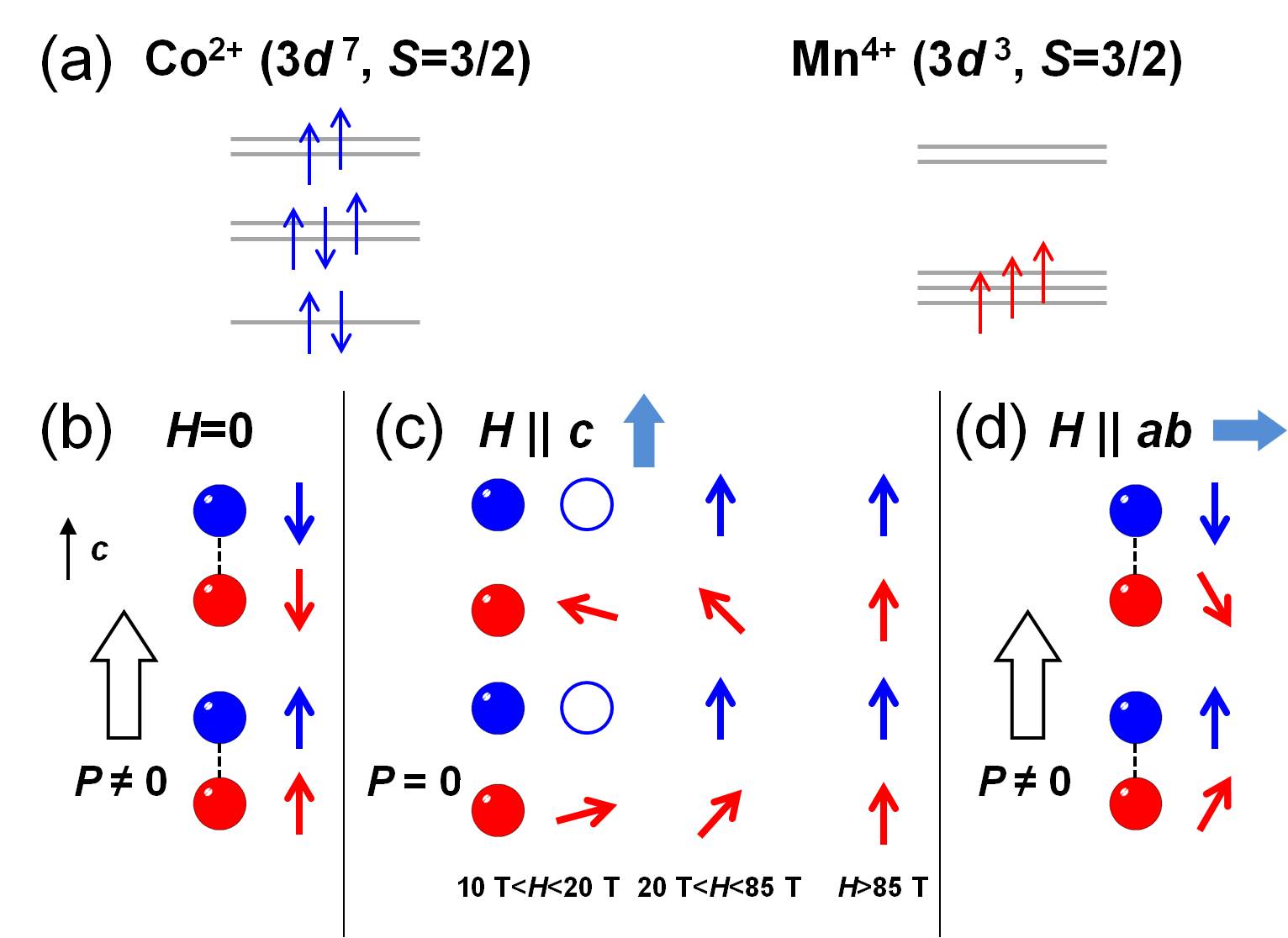}
\caption{ 
(a) Crystal-field level occupations for Co$^{2+}$ in a trigonal prismatic environment (left) and Mn$^{4+}$ in an octahedral environment (right).
Schematic diagrams of spin configurations and ionic displacements at (b) zero field, (c) magnetic field along the $c$-axis in different field values, and (d) perpendicular to the $c$-axis.
Blue and red filled circles represent Co$^{2+}$ and Mn$^{4+}$ ions, respectively.
Blue open circles represent the unknown Co$^{2+}$ spin configuration between 10 and 20~T.
Dashed lines indicate distance between neighboring ions that have been shortened by magnetostriction, which is thought to be the mechanism that leads to electric polarization \cite{Choi08}.
}
\label{model}
\end{figure}

However, further model refinement for CCMO will require understanding how Ising-like the Co$^{2+}$ and Mn$^{4+}$ spins really behave, as well as the spin states, which are currently controversial.
Jo $et$ $al.$ \cite{Jo09} reported NPD and magnetization measurement of single crystals up to 11 and 33~T, respectively.
At low $T$, NPD data were fit to an $\uparrow \uparrow \downarrow \downarrow$ state at zero field, and an $\uparrow \uparrow \uparrow \downarrow$ state at 11~T with a 3~$\mu_B$/formula unit (f.u.) magnetization plateau \cite{Jo09}.
Another quasi-plateau forms above 20~T with 4~$\mu_B$/f.u., which they tentatively attribute to complete saturation.
Therefore, they identify Co$^{2+}$ and Mn$^{4+}$ ions being $S$~=~1/2 and 3/2, respectively.
However, an X-ray absorption spectroscopy (XAS) study at room temperature \cite{Wu09} and Curie-Weiss fits to the susceptibility between 75 and 300~K in compounds with similar $x$ values \cite{Zubkov01} were more consistent with both the Co$^{2+}$ and Mn$^{4+}$ ions having the $S$~=~3/2 spin state (Fig.~\ref{model}(a)).
Both interpretations are consistent with the emergent $P$ at low temperatures, since the breaking of SIS is not related to the spin amplitude.
Flint $et~al.$ combine these two scenarios in a model based on $S$~=~1/2 Co$^{2+}$ ions at $H$~=~0 and low $T$, with a magnetic field-driven spin-state crossover to $S$~=~3/2 in applied magnetic fields \cite{Flint10}.

In this Communication, we determine the spin states in CCMO by studying magnetization, electric polarization, magnetostriction, and magnetocaloric effect (MCE) up to 92~T, which is above magnetic saturation.
Based on our measurements, we find that both Co$^{2+}$ and Mn$^{4+}$ magnetic ions have $S$~=~3/2 at all magnetic fields and we propose a different spin configuration at high magnetic fields from previous works.
The new model provides a different understanding of how the evolving magnetic order destroys electric polarization, involving spin flops of quasi-isotropic Mn$^{4+}$ spins.

Single crystals of Ca$_3$Co$_{2-x}$Mn$_{x}$O$_6$ with $x \sim$~0.97 were synthesized as in previous works \cite{Choi08, Jo09} where $x$ was identified from magnetic susceptibility measurements \cite{Kiryukhin09}.
High magnetic field measurements were performed using various magnets driven by capacitors, a generator, or both (the 100~T magnet) at the NHMFL pulsed-field facility at LANL.
Magnetization was measured by using an induction magnetometry technique \cite{Detwiler00} up to 92~T.
The pulsed-field magnetization values were calibrated against measurements in a 14~T DC magnet using a vibrating sample magnetometer.
A systematic error bar in the pulsed-field magnetization values at 85~T of $\pm$0.5 $\mu_{B}$/f.u. results from the uncertainty created by hysteresis and sweep-rate dependences when compared to DC measurements.
Magnetostriction was measured in the 100~T hybrid pulse magnet along the $c$-axis using an optical fiber with a Bragg-grating \cite{Daou09,Jaime12}.
MCE was measured in the generator-driven 60~T shaped-pulse magnet by reading the temperature sensor attached to the sample while sweeping the magnetic field with the sample immersed in superfluid $^4$He.
This thermal setup was chosen because the alternate option of measuring in vacuum resulted in a semi-adiabatic thermal situation where the temperature relaxations occurred on the same time scale as $H$-induced temperature changes, making analysis difficult \cite{ZapfRMP}.
Electric polarization was measured in the 65~T capacitor-driven magnet by recording the magnetoelectric current during a magnetic field pulse and integrating it in time (see SI, Fig.~S2) \cite{Zapf10}.
Prior to the measurement, samples were poled by cooling from 40~K to 1.5~K in a static poling electric field of 645~kV/m.

\begin{figure}
\includegraphics[width=0.42\textwidth]{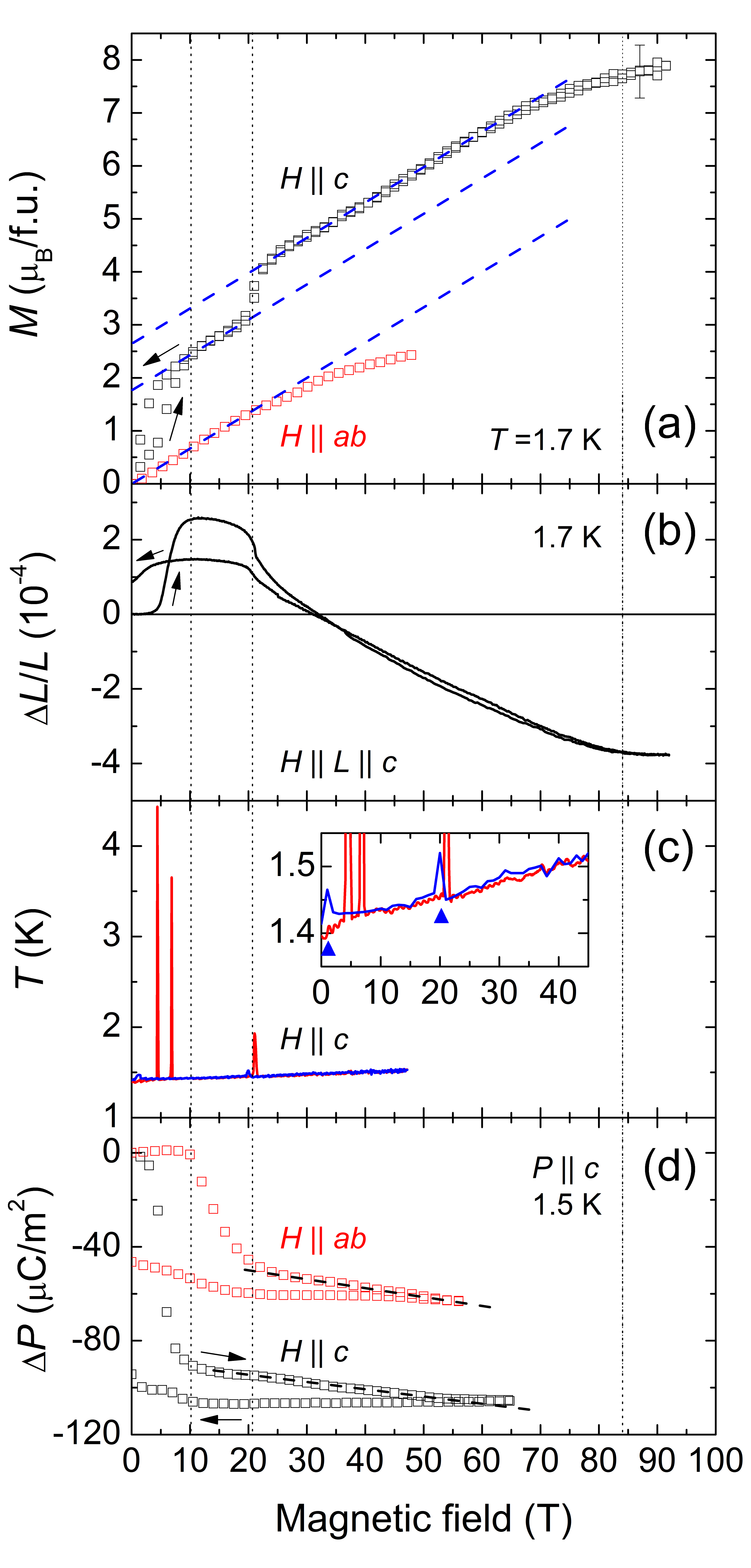}
\caption{(a) Magnetization, (b) magnetostriction, (c) magnetocaloric effect, and (d) change of electric polarization ($\Delta$$P$($H$)~$\equiv$~$P$($H$)-$P$($H$~=~0)) as a function of magnetic field of CCMO ($x \sim $0.97).
(a) $M$($H$) was measured along the $c$-axis (black squares) and $ab$-plane (red squares).
Blue dashed lines are guides to the eye that show identical $M$($H$) slopes.
The error bar at 85~T is $\pm$0.5~$\mu_B$/f.u..
(c) Red and blue curves denote up and down sweep measurements, respectively.
Inset shows the expanded region near base temperature indicating the $T$ jumps in the down sweep (blue triangles).
(d) $\Delta$$P$($H$) was measured along the $c$-axis with different magnetic field directions.
The dotted and dashed lines are guides to the eye.
}
\label{M}
\end{figure}

Fig.~\ref{M}(a) shows the $M$($H$) curves with the magnetic field along different crystallographic directions.
For $H \parallel c$, $M$($H$) shows two plateau-like features, similar to those seen previously in DC field measurements up to 33~T \cite{Jo09}.
There is a small discrepancy between the value of the plateau between 10 and 20~T, which is 3~$\mu_B$/f.u. at 15~T in the DC data and 2.7~$\mu_B$/f.u. at 15~T in our pulsed-field data.
However, the DC $M$($H$) data actually shows different values of this plateau for positive and negative sweeps and the pulsed-field data agrees with the 2.5~$\mu_B$/f.u. value seen for negative DC field sweeps.
A second quasi-plateau occurs in $M$($H$) between 20 and 30~T with an onset value of 4~$\mu_B$/f.u..
The important observation from our data is that this quasi-plateau is not the final saturation, but rather $M$($H$) continues to increase above 33~T and reaches saturation magnetization $M_{sat}$~=~7.7$\pm$0.5~$\mu_{B}$/f.u..
The final saturation magnetic field ($H_{\rm sat}$) can be most accurately determined from the magnetostriction data (Fig.~\ref{M}(d)), which shows a change of slope approaching saturation at 85~T.
Magnetostriction also shows features at similar fields to $M$($H$) although $\Delta$$L$($H$)/$L$ is non-monotonic (Fig.~\ref{M}(b)).

This saturation value of $M$ requires both the Co$^{2+}$ and Mn$^{4+}$ ions to be in the high spin state ($S$~=~3/2) with an additional orbital contribution.
The orbital contribution of 1.7$\pm0.5~\mu_B$ per Co$^{2+}$ is consistent with $ab$-$initio$ calculations \cite{Wu09}, while Mn$^{4+}$ (3$d$$^3$) is in an octahedral symmetry that cannot have an orbital contribution.
We also measured $M$($H$) for $H \parallel ab$, shown in Fig.~\ref{M}(a).
In this configuration, $M$($H$) increases linearly up to 19~T and then the slope begins to decrease.
No hysteresis was observed for $H  \parallel  ab$.
There exists a common background linear slope in $M$($H$) for both $H \parallel c$ and $H \parallel ab$, which will be discussed later.

The existence of a magnetic-field-induced low-to-high spin state transition (LHST) of the Co$^{2+}$ spins can be checked by the MCE measurement.
In the vicinity of a LHST, multiple spin states become available which should increase the entropy of the spins, and in turn reduce the entropy and the temperature of the lattice via conservation of entropy.

MCE up to 50~T is shown in Fig.~\ref{M}(c) in the limit where the thermal relaxation time is shorter than the experiment time.
Thus, we expect to see jumps in the temperature at phase transitions followed by a rapid relaxation, and the direction of the spike tells us if the spin entropy increases (lattice $T$ decreases) or decreases (lattice $T$ increases) \cite{ZapfRMP}.
The data included in Fig.~\ref{M}(c) shows three upward jumps at 4.5, 6.9, and 21~T during the up sweep.
The 21~T spikes correspond to features in $M$($H$) and $\Delta$$L$($H$)/$L$, while the 4.5 and 6.9~T spike only appears for certain sweep rates \cite{Kim13}.
During the down sweep of the magnetic field, the temperature also shows upward spikes at 1.5 and 20~T and with greatly reduced amplitude ($\Delta$$T$~$<$~0.1~K, inset in Fig.~\ref{M}(c)).
This hysteresis in the amplitude and field between up and down sweeps indicates that there are both reversible and irreversible components in the phase transition.
Thus, we see no evidence of a LHST in the MCE at any of the sharp transitions up to 25~T, and for magnetic fields higher than that, the value of $M$ requires the Co$^{2+}$ spin to already exceed $S$~=~1/2.
Finally, the sign of $\Delta$$L$/$L$ is important to LHST.
The high spin Co$^{2+}$ ($S$~=~3/2) ion is significantly larger than low spin Co$^{2+}$ ($S$~=~1/2).
In CCMO, $\Delta$$L$/$L$ decreases for $H$~$>$~20~T with a relative magnitude of 10$^{-4}$, which makes a LHST in that magnetic field range very unlikely.

Fig.~\ref{M}(d) shows the change of $c$-axis electric polarization in CCMO relative to the value at $H$~=~0 measured for $H \parallel ab$ and $H \parallel c$.
For $H \parallel c$, the sharp drop in $P$ below 10~T is consistent with previous DC measurements \cite{Jo09}, and with features in $M$($H$), $\Delta$$L$($H$)/$L$, and MCE (Fig.~\ref{M}).
However, above 10~T, the pulsed-field data shows changes that were not resolved in DC field measurement.
This difference is partially due to the fact that fast sweep rates of the magnetic field inherently increase resolution of electric polarization measurements (see Fig.~S2), but also because the change in $P$ for $H$~$>$~10~T becomes more pronounced above 20~T.
The $\Delta$$P$($H$) slope above 20~T for $H \parallel c$ is similar to that of $\Delta$$P$ above 25~T in the $H \parallel ab$ configuration.

Thus, a picture for CCMO emerges in which all spins are $S$~=~3/2 both at low and high magnetic fields (Fig.~\ref{model}(a)).
This observation agrees with room-temperature XAS and high-temperature Curie-Weiss fits \cite{Wu09,Zubkov01}.
However, NPD studies have suggested $S$~=~1/2 for Co$^{2+}$ and $S$~=~3/2 for Mn$^{4+}$ ordered moments in an $\uparrow \uparrow \downarrow \downarrow$ configuration at $H$=0 and $\uparrow \uparrow \uparrow \downarrow$ for 11~T \cite{Choi08,Jo09}.
We emphasize that NPD experiment determines the size of the $ordered$ moment, not the total moment.
Reduction in the ordered moment can be accounted for by fluctuations due to frustration or by disorder due to Co$^{2+}$-Mn$^{4+}$ site interchange, and also by (possibly disordered) long-wavelength modulations as were observed in Ca$_3$Co$_2$O$_6$ \cite{Moyoshi11,Agrestini08}.
Alternate interpretations of the NPD data can allow for the reduced ordered moment to be on the Mn$^{4+}$ instead of the Co$^{2+}$ site, or shared between the two.

Besides the $S$~=~3/2 spin amplitude, our data also shows evidence for quasi-isotropic Mn$^{4+}$ spins.
In past models of CCMO, both the Co$^{2+}$ and Mn$^{4+}$ spins were treated as effectively Ising-like and oriented along the $c$-axis \cite{Yao09,Flint10}, and the assumption was that the Mn$^{4+}$ spin was always strongly clamped to the Ising-like Co$^{2+}$ spin.
While collinear spins are consistent with NPD data at $H$~=~0 showing $\uparrow \uparrow \downarrow \downarrow$ order \cite{Jo09}, in applied magnetic fields our $M$($H$), $\Delta$$L$($H$)/$L$, and $\Delta$$P$($H$) data show extended regions with linear slopes that strongly point to canting of quasi-isotropic spins.
The guide lines shown in Fig.~\ref{M}(a) highlight the background linear slope in $M$($H$) that is the same for $H \parallel ab$ and $H \parallel c$, and extends from 25 to almost 70~T for $H \parallel c$.
A similar linear slope in $M$($H$) is seen in DC $M$($H$) measurements within the plateaus \cite{Jo09}.
Between 10 and 25~T, this linear slope coexists with step-like behavior that is more characteristic of Ising spins.
Therefore we suggest that one species is predominantly Ising-like and the other is quasi-isotropic with a significant spin flop into the $ab$-plane.
Since the Mn$^{4+}$ ion is in an octahedral site symmetry with one  electron in each $t$$_{\rm2g}$ level, its orbital moment is quenched and thus it is likely the quasi-isotropic species, whereas Co$^{2+}$ ion with a trigonal prismatic site symmetry is expected to be Ising-like (Fig.~\ref{model}(a)).

Our data is inconsistent with the 11~T collinear $\uparrow$(Mn$^{4+}$)-$\uparrow$(Co$^{2+}$)-$\uparrow$(Mn$^{4+}$)-$\downarrow$(Co$^{2+}$) state that was previously proposed as one interpretration of 11~T NPD and magnetization data \cite{Jo09}.
The continued linear evolution of $M$($H$) to fields beyond 11~T does not allow the Mn$^{4+}$ spins to be polarized by 11~T.
The reverse state, $\uparrow$(Co$^{2+}$)-$\uparrow$(Mn$^{4+}$)-$\uparrow$(Co$^{2+}$)-$\downarrow$(Mn$^{4+}$), is also inconsistent with our results because it would produce a magnetization that is larger than what we observe, given Co$^{2+}$ $S$~=~3/2 moments with 1.7~$\mu_B$ orbital contribution.
In order to account for our $M$($H$) data, the Mn$^{4+}$ moments must flop into the $ab$-plane at low fields and then subsequently cant along $H \parallel c$ as $H$ increases (Fig.~\ref{model}(c)), as is typical for quasi-isotropic antiferromagnets.
We find that NPD work at 11~T \cite{Jo09} (of which some of us are co-authors) does not exclude the scenario of flopped Mn$^{4+}$ spins.
Further elastic neutron diffraction measurements in applied magnetic fields on single crystals should be able to resolve the details of the Mn$^{4+}$ moment ordering and the spin structure in the first magnetization plateau (10~T~$<$~$H$~$<$~20~T).

We note that the plateau-like behavior in $M$($H$) and sweep-rate-dependent steps \cite{Kim13} measured along the $c$-axis stops by 25~T leaving only a near-linear evolution to saturation.
From this we posit that the Ising Co$^{2+}$ spins dominate the behavior up to 25~T, progressing through a series of different ordered phases as is typical for frustrated Ising spins, but then saturate by 25~T leaving the quasi-isotropic Mn$^{4+}$ spins to continue canting until their saturation by 85~T as is sketched in spin structures in Fig.~\ref{model}(c).
The energy scale of the effective Mn$^{4+}$-Mn$^{4+}$ exchange interaction is quantified by the linear slope in $M$($H$) as $\sim $10~K.
The saturation at 85~T is the result of overcoming the Mn$^{4+}$-Mn$^{4+}$ exchange, but in the presence of the effective molecular field of the saturated Co$^{2+}$ spins.
The magnetostriction also changes from increasing steplike with magnetic field from 0 to 25~T to decreasing continuously with magnetic field above 25~T (Fig.~\ref{M}(b)).
This implies that the magnetic forces due to effective Co$^{2+}$-Co$^{2+}$ magnetic exchange, which contribute to the magnetostriction below 25~T, have an opposite effect on the $c$-axis lattice constant than those from effective exchange bonds connected to Mn$^{4+}$ (Mn$^{4+}$-Mn$^{4+}$ and Mn$^{4+}$-Co$^{2+}$) that control the magnetostriction above 25~T.

The above conclusions call for a different interpretation of magnetically-controlled electric polarization in CCMO.
Previously, the magnetic field-induced suppression of $P$ was attributed to the transition from a collinear $\uparrow \uparrow \downarrow \downarrow$ to another collinear $\uparrow \uparrow \uparrow \downarrow$ state.
However, this does not explain the simultaneous occurrence of linear slope in $M$($H$) and suppression of $P$($H$) above 10~T (Figs.~\ref{M}(a) and \ref{M}(d)).
Instead, we find that a non-collinear spin structure that arise from spin flop of Mn$^{4+}$ spins into the $ab$-plane, well explains both features.
The continuous evolution of $\Delta$$P$($H$) above 20~T with a common slope for both directions of the magnetic field may be due to (1) a configuration of Mn$^{4+}$ spin components in the $ab$-plane that allows for broken SIS, (2) local regions of electric polarization that persist to high magnetic fields due to Mn$^{4+}$-Co$^{2+}$ site interchange and off-stoichiometry, and (3) dynamic effects due to the magnetic sweep rate in pulsed magnets \cite{Kim13}.
Interestingly, in the $H \parallel ab$ configuration, $\Delta$$P$($H$) is flat up to 10~T and then decreases with applied magnetic field.
Finally, we note that the $P$ value when $H$ is along the $ab$-plane is larger than that of the $H \parallel c$ case by $\sim $50~$\mu$C/m$^{2}$ at 60~T which is suggestive of a robust magnetic structure with broken SIS.
In this configuration, one can postulate that the $\uparrow \uparrow \downarrow \downarrow$ structure is preserved along the $c$-axis since $H \parallel ab$ cants only the Mn$^{4+}$ spins and allows for spin components along the $c$-axis as illustrated in Fig.~\ref{model}(d).
Further neutron diffraction work on single crystal is required to understand this behavior.


In conclusion, the high magnetic field experiments show that both Co$^{2+}$ and Mn$^{4+}$ moments are in the high spin state with $S$~=~3/2, and no LHST is seen in applied magnetic fields.
We find regions of continuous evolution of the magnetization that strongly support canting of Mn$^{4+}$ ($S$~=~3/2) quasi-isotropic spins.
When magnetic field is applied along the $c$-axis, the Mn$^{4+}$ moments thus have a spin-flop into the $ab$-plane at low fields followed by subsequent canting towards the $c$-axis.
Sharp steps and hysteresis that are characteristic of frustrated Ising spins are observed in the magnetization, electric polarization, magnetostriction, and MCE up to 25~T, due to the evolution of frustrated Ising Co$^{2+}$ spins that saturate at 25~T, leaving the quasi-isotropic Mn$^{4+}$ spins to cant continuously towards saturation at 85~T.
We observe an electric polarization extending to higher magnetic fields (at least up to 60~T) than previously observed (10~T, \cite{Jo09}), which indicates a remanent SIS breaking for the high-field magnetically-ordered phases.
CCMO shows many hallmarks of ANNNI physics for $H$~$<$~25~T as suggested previously for CCMO \cite{Choi08,Jo09,Yao09,Lancaster09,Flint10} and for the related compound Ca$_3$Co$_2$O$_6$ \cite{Agrestini08,Moyoshi11,Kamiya12b}.
However, a model as well as recent experiments for CCMO will need to take into account the interaction of Co$^{2+}$ Ising spins with Mn$^{4+}$ quasi-isotropic spins \cite{Jo09,Flint10,Yao09,Wu09,Zhang09,Sheng13,Ruan13a,Ruan13b}.
The data quantifies several key parameters necessary for modeling: (1) the spin amplitudes of Co$^{2+}$ and Mn$^{4+}$ ions ($S$~=~3/2), (2) the respective saturation fields of the Co$^{2+}$ and Mn$^{4+}$ spins (25 and 85~T), and (3) the Mn$^{4+}$-Mn$^{4+}$ exchange interaction from the slope of the magnetization ($\sim$~10~K).

The NHMFL facility is funded through the U.S. NSF Cooperative Grant No. DMR-1157490, the DOE, and the State of Florida.
This work is supported by the DOE BES project "Science at 100 tesla", the DOE's Laboratory-Directed Research, and Development program.
The work at Rutgers was supported by the DOE Award No. DE-FG02-07ER46328.
We acknowledge discussions with R.~Flint, P.~Chandra, and G.~Pascut.

\bibliography{CCMO-140115}

\begin{thebibliography}{43}
\bibitem [{Cou()}]{Coupling}%
  \BibitemOpen
  \href@noop {} {}\bibinfo {note} {A partial list some of the stronger
  couplings between $M$ and $P$ include some of the rare earth manganites
  \cite{Kimura03,Hur04}, and hexaferrites \cite{Kimura12,Chun12}. CoCr$_2$O$_4$
  is an example of hysteretic coupling between net $M$ and $P$, although much
  weaker than in Ca$_3$Co$_{2-x}$Mn$_x$O$_6$ \cite{Yamasaki06}.}\BibitemShut
  {Stop}%
\bibitem [{Kim()}]{Kim13}%
  \BibitemOpen
  \href@noop {} {}\bibinfo {note} {We have observed magnetic field sweep-rate
  dependence in the magnitudes of $M$($H$) and $\Delta$$L$($H$)/$L$ at 1.5~K,
  and we see additional small steps in $\Delta L(H)$ that between 4 and 7~T for
  sweep rates between 74 and 1400~T/s. These could be analgous to the
  sweep-rate dependence in Ca$_3$Co$_2$O$_6$, although in that sample the sweep
  rates are much slower. This data will be published separately.}\BibitemShut
  {Stop}%
\end{thebibliography}

\end{document}